\documentclass{ws-spin}
\usepackage{multicol}
\usepackage{hyperref}
\usepackage{graphicx}
\usepackage{epstopdf}
\usepackage{amssymb, amsmath}
\usepackage[usenames]{color}
\usepackage{bm}
\usepackage{multirow}
\usepackage{dcolumn}
\usepackage{float}
\hypersetup{
	colorlinks=true,
	linkcolor=blue,
	filecolor=gray,      
	urlcolor=blue,
	citecolor=blue,
}
\begin{document}

\title{Quantum Spin Hall Insulators in Tin Films: Beyond Stanene}
\author{Jiaheng Li}
\address{State Key Laboratory of Low Dimensional Quantum Physics and Department of Physics, Tsinghua University, Beijing, 100084, China\\
}
\author{Yong Xu}
\address{State Key Laboratory of Low Dimensional Quantum Physics and Department of Physics, Tsinghua University, Beijing, 100084, China\\
		RIKEN Center for Emergent Matter Science (CEMS), Wako, Saitama 351-0198, Japan \\
\email{yongxu@mail.tsinghua.edu.cn}
}

\maketitle

\begin{abstract}
Large-gap quantum spin Hall (QSH) insulators were previously predicted in stanene and its derivatives. Beyond stanene that is the thinnest $\alpha$-Sn(111) film, we propose to explore QSH insulators in $\alpha$-Sn films with different crystallographic orientations. Our first-principles calculations reveal that the thickness-dependent band gap of $\alpha$-Sn (100) and (110) films does not show a monotonic decrease as typically expected by quantum confinement, but displays an oscillating change behavior, an indicative of topological quantum phase transition. While these films are normal insulators in the ultrathin limit, the QSH phase emerges above a critical film thickness of around 10 layers. Remarkably, the QSH insulators are obtainable within a wide thickness range and their energy gaps are sizable (even $>$0.1 eV), which facilitates experimental realization of the high-temperature QSH effect.
\end{abstract}


\begin{multicols}{2}
\section{Introduction}

Great research effort has been devoted to explore exotic topological quantum physics and materials since the discovery of topological insulators (TIs)\cite{hasan2010,qi2011}. Among them, the quantum spin Hall (QSH) effect\cite{kane2005,bernevig2006,konig2007} and the quantum anomalous Hall (QAH) effect\cite{haldane1988,liu2008,yu2010,chang2013,liu2016} are of special interest, which are able to give quantum-Hall-like edge states without applying magnetic field and thus are promising for low-power electronics\cite{hasan2010,qi2011}. The QSH and QAH effects were first experimentally observed in HgTe/CdTe quantum well\cite{bernevig2006,konig2007} and magnetically doped TI thin films\cite{liu2008,yu2010,chang2013,liu2016}, respectively. However, these material systems are difficult to fabricate and control, and the quantized Hall conductance was only observable at very low temperatures. The low working temperature is mainly limited by the small topological energy gap of the bulk. In this context, the key issue is to search for large-gap QSH and QAH insulators in simple material systems.

Recently, numerous two-dimensional (2D) topological materials have been theoretically predicted\cite{murakami2006,liu2008_2,liuzheng2011,liu2011prl,xu2013,tang2014,qian2014,li2019,wang2019,olsen2019}, and few experimentally reported\cite{knez2011,reis2017,fei2017,tang2017,wu2018science,deng2019,liu2019,ge2019}. Among these candidate materials, stanene is of great interest and much research progress has been made on this intriguing material\cite{molle2017}. Stanene is a monolayer of tin (Sn) crystallized in a honeycomb lattice, which can be viewed as the Sn analogy of graphene\cite{xu2013}. Similar as graphene, stanene is a QSH insulator described by the Kane-Mele model\cite{kane2005,xu2013,liu2011}. Differently, its bulk energy gap is about five orders of magnitude larger than that of graphene, due to the much stronger spin-orbit coupling (SOC) of stanene caused by its heavier element and structural buckling\cite{molle2017}. Remarkably, by saturating the $p_z$ orbitals of stanene with variant chemical groups, a series of new 2D materials are derived, which are not only chemically stable but also topologically nontrivial\cite{xu2013}. In contrast to bare stanene, the stanene derivatives are characterized by a $s$-$p$ band inversion at the $\Gamma$ point described by the Bernevig-Hughes-Zhang (BHZ) model\cite{bernevig2006}. Also they can be effectively described by another multi-orbital ($p_{xy}$) honeycomb model, which was theoretically proposed to host novel topological physics\cite{wu2007,zhang2014}. In addition to the large-gap QSH states, stanene and its derivatives have been theoretically predicted to display many other exotic topological quantum physics, such as the near-room temperature QAH effect\cite{wu2014}, enhanced thermoelectric performance\cite{xu2014}, and 2D topological superconductivity\cite{wang2014}.

On the experimental side, monolayer stanene has been successfully fabricated and experimentally confirmed for the first time on the Bi$_2$Te$_3$(111) substrate\cite{zhu2015}. Unfortunately, since the valence bands of stanene are pinned with the conduction bands of Bi$_2$Te$_3$, the interface becomes metallic. Further theoretical work suggested some substrates for realizing the 2D TI phase in decorated stanene\cite{xu2015prb}. In the following experiments, an insulating monolayer stanene was fabricated on PbTe(111)\cite{zang2018}, where the surface $p_z$ orbitals of stanene were spontaneously saturated in experiments, possibly by hydrogen. The decorated stanene, however, was identified to be topologically trivial due to the small lattice constant. Very recently, ultraflat stanene was experimentally discovered on Cu(111), which has an extremely large lattice constant of 5.1 \AA \  and thus displays a topological band inversion\cite{deng2018}. The SOC-induced band gap and topologically derived edge states were experimentally detected, in agreement with theoretical predictions. One shortage is that the metallic substrate is not suitable for observing the quantized Hall conductance. Further research effort is still needed to realize the QSH effect in stanene.

Another promising research direction is beyond monolayer stanene that is the thinnest (111) film of $\alpha$-Sn to study few-layer stanene (i.e., $\alpha$-Sn(111) films)\cite{chou2014,liao2018}. According to previous calculations, few-layer stanenes are mostly metallic, and they can be QSH insulators when the thickness is less than 4 layers\cite{chou2014}. Recent experiments explored the size effect of few-layer stanenes, and unexpectedly discovered 2D superconductivity in them\cite{liao2018}, although the bulk $\alpha$-Sn is well known to be non-superconducting. The coexistence of topology and superconductivity opens new opportunities to explore topological superconductivity in the stanene material system. Very recently, type-II Ising superconductivity, a new kind of 2D superconductivity protected against magnetic field by SOC without breaking inversion symmetry, was theoretically predicted in 2D materials with rotational symmetries (including stanene)\cite{wang2019prl} and experimentally discovered in few-layer stanene\cite{falson2019} and PdTe$_2$\cite{liuyi2019}.

In this work, we investigate electronic and topological properties of $\alpha$-Sn films with different crystallographic orientations by first-principles calculations. In contrast to usual cases that the band gap monotonically decreases with increasing film thickness, we find that the thickness-dependent band gap of $\alpha$-Sn (100) and (110) films displays an oscillating  behavior. These films are normal insulators in the ultrathin limit, and a topological quantum transition into the QSH phase occurs when the film thickness increases beyond a critical value about 10 layers. Consequently, large-gap QSH insulators emerge in the $\alpha$-Sn films with a wide thickness range. Compared to the previous proposals of stanene\cite{xu2013,xu2015prb}, the present proposal of QSH insulators does not reply on strict conditions of substrate and film thickness, which is very likely to be realized experimentally.

\section{Computational Details}

First-principles calculations were performed by the Vienna $ab$ $initio$ simulation package\cite{kresse1996} in the framework of density functional theory (DFT). The energy cutoff of plane-wave basis was set at 400 eV. The projector augmented wave method was adopted to describe the electron-ion interactions. Two types of exchange-correlation functionals were used, including the  Perdew-Burke-Ernzerhof (PBE) type in the generalized gradient approximation\cite{perdew1996} and the Heyd-Scuseria-Ernzerhof (HSE) hybrid functional\cite{krukau2006}. Compared to DFT-PBE, DFT-HSE typically provides a better description on the band gap, but is computationally much more expensive. Thus, DFT-HSE was only applied to compute electronic band structures, and only thin films were considered due to the limitation of computational ability.

Sn films were constructed by cutting atomic layers from the bulk $\alpha$-Sn, and surface atoms of Sn films were decorated by hydrogen atoms for removing dangling bonds. In practice, the dangling bonds might be saturated differently, for instance, by surface reconstruction or by other kinds of chemical decoration. They are expected to give similar results, since the saturation group has minor contribution to the electronic states near the Fermi level. The slab model together with a vacuum region of over 15 \AA \ was applied to model Sn films. Calculations of 3D $\alpha$-Sn bulk used an experimental lattice constant of 6.43 \AA. The structural relaxation on both lattice constants and atomic positions was performed with a force criterion of 0.01 eV/\AA \ for thin-film calculations. The spin-orbit coupling (SOC) was included self-consistently in band-structure calculations. The Monkhorst-Pack $k$-point mesh of 6$\times$6$\times$6 was adopted for bulk $\alpha$-Sn, 7$\times$7$\times$1 for Sn(100) films, and 5$\times$7$\times$1 for Sn(110) films. Topological invariant $Z_2$, Wannier charge centers and edge-state calculations were performed by the WannierTools package\cite{wu2018cpc}, using tight-binding Hamiltonians based on maximally localized Wannier functions constructed from DFT calculations.

\section{Results and Discussion}

\begin{figure*}[htbp]
	\includegraphics[width=0.7\linewidth]{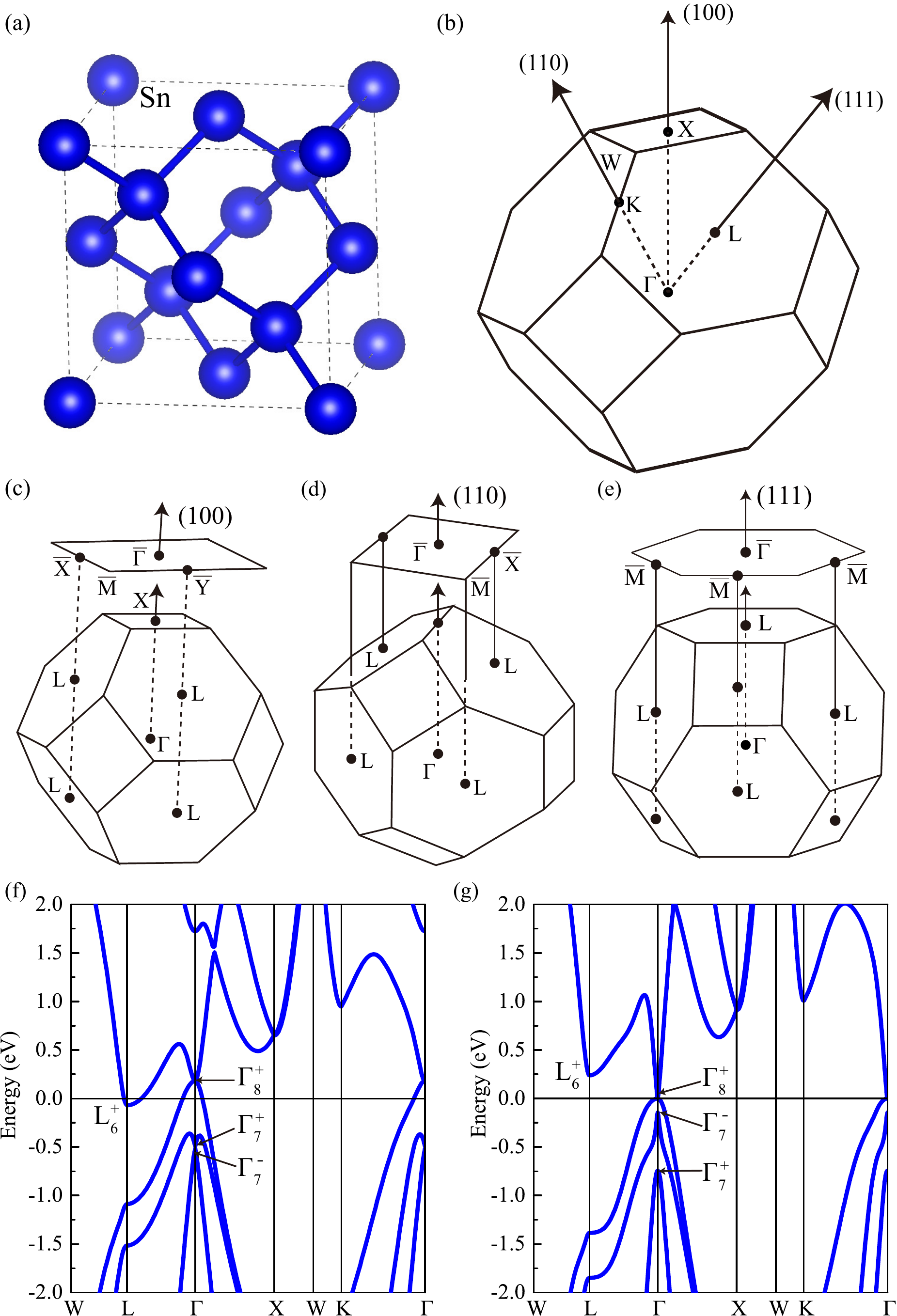}
	\centering
	\caption{Physical properties of $\alpha$-Sn bulk. (a) The atomic structure and (b) Brillouin zone. (c,d,e) The surface Brillouin zone of (100), (110) and (111) surfaces. (f,g)  Band structures of $\alpha$-Sn bulk calculated by (f) DFT-PBE and (g) DFT-HSE.} \label{fig1} 
\end{figure*} 

Tin (Sn) has two major allotropes in nature, including gray tin ($\alpha$-Sn) with a diamond cubic crystal structure (Fig. \ref{fig1}(a)) and white tin ($\beta$-Sn) with a body-centered tetragonal structure. The bulk $\alpha$ and $\beta$ phases are natively stable below and above $\sim$13.2$^{\circ}$C, respectively\cite{madelung2012}. In contrast, the $\alpha$-$\beta$ structural phase transition could occur at much higher temperatures in films grown on substrates. For example,  $\alpha$-Sn(111) thin films fabricated on InSb($\bar{1}$$\bar{1}$$\bar{1}$) were experimentally found to display a structural transition from the $\alpha$ phase directly into the liquid phase when heated at $170^{\circ}$C\cite{kasukabe1988}. 

\begin{figure*}[htbp]
	\centering
	\includegraphics[width=0.6\linewidth]{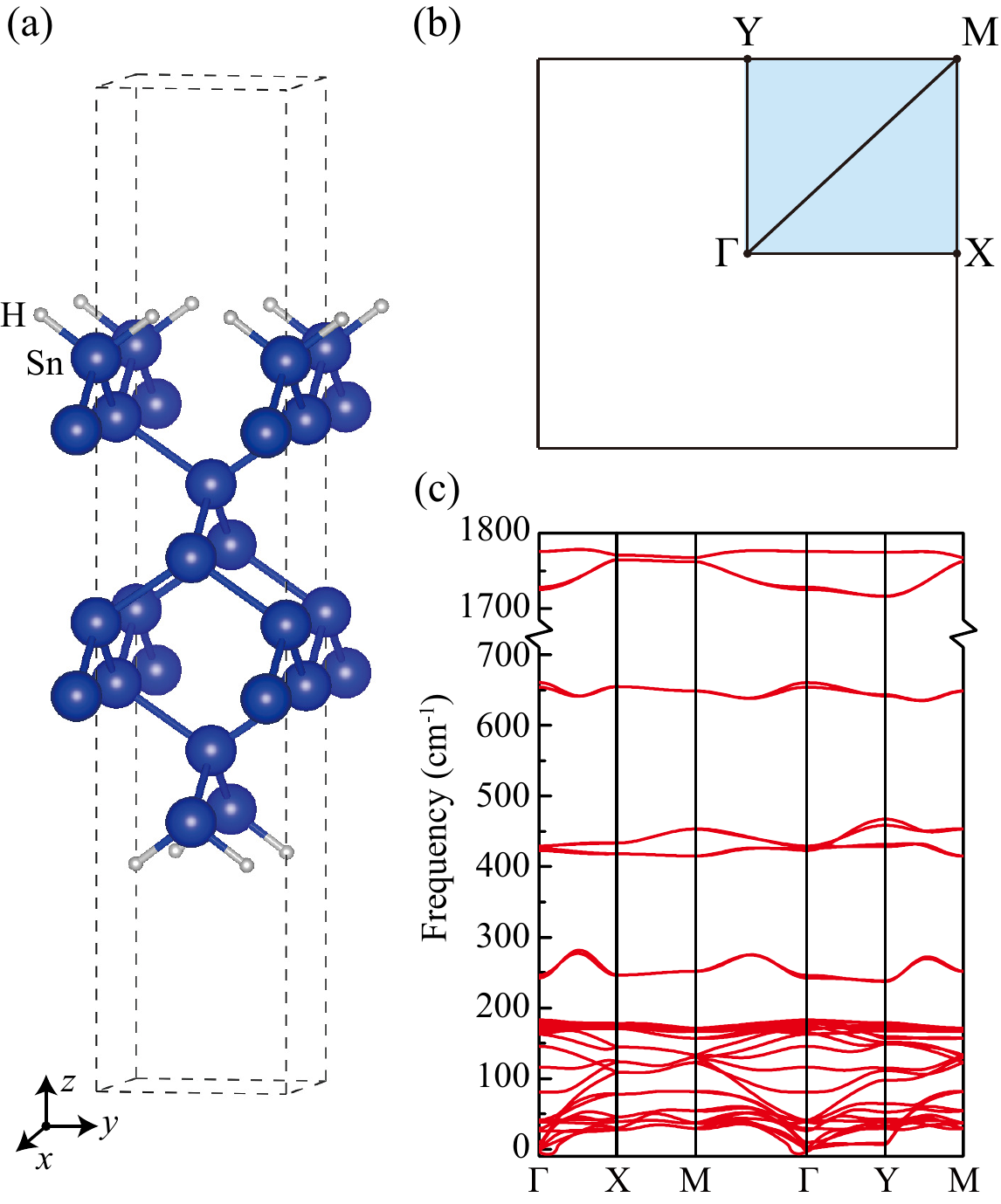}
	\caption{Physical properties of $\alpha$-Sn (100) films. (a) The optimized atomic structure ($a=4.75$ \AA, $b=4.69$ \AA), (b) 2D surface Brillouin zone and (c) phonon dispersion of an 8-layer $\alpha$-Sn (100) film. Sn and H atoms are denoted by blue and gray balls, respectively.} \label{fig2} 
\end{figure*} 

While $\beta$-Sn is a normal metal, $\alpha$-Sn is a zero-gap semiconductor, which is well known to have an inverted band gap (called ``negative energy gap'') at $\Gamma$, similar as HgTe and HgSe\cite{madelung2012}. Figures \ref{fig1}(f) and (g) present band structures of $\alpha$-Sn bulk calculated by DFT-PBE and DFT-HSE, respectively. $\alpha$-Sn has both spacial inversion symmetry and time reversal symmetry. Thus every band is spin degenerate. Two kinds of Bloch states at $\Gamma$ are relevant in our discussion, including the $s$-like $\Gamma_7^-$ state and the $p$-like $\Gamma_8^+$ and $\Gamma_7^+$. They are characterized by opposite parities, which do not hybridize with each other. $\Gamma_8^+$ and $\Gamma_7^+$ are originally degenerate when excluding the SOC and get split when including the SOC. Normally, the anti-bonding $\Gamma_7^-$ state is located above the bonding $\Gamma_8^+$ state, as found in the diamond structure of Si and Ge. Distinctly, $\Gamma_7^-$ shifts below $\Gamma_8^+$ in $\alpha$-Sn, giving an inverted band order. This band inversion is topologically nontrivial, according to the parity criterion\cite{fu2007}. However, the highest valence band and the lowest conduction band at $\Gamma$ are both contributed by the fourfold degenerate $\Gamma_8^+$ state, implying a zero band gap. The fourfold degeneracy is protected by cubic symmetry and time reversal symmetry. Breaking either symmetry could open the band gap, leading to topologically nontrivial states. For instance, $\alpha$-Sn grown on InSb(001) becomes a 3D TI showing 2D gapless topological surface states, which is explained by strain and/or quantum confinement effects\cite{barfuss2013,ohtsubo2013}. $\alpha$-Sn can be also driven into the 3D topological Dirac semimetal phase by applying a uniaxial tensile strain\cite{huang2017,xu2017}, which can be tuned into the 3D topological Weyl semimetal phase if further breaking time reversal symmetry by magnetism or magnetic field.

Moreover, special attention should be paid on the $L_6^+$ state, which is located near the Fermi level and plays an important role in determining properties of thin films. In DFT-PBE, the ordering of states by increasing energy is $\Gamma_7^-$-$\Gamma_7^+$-$L_6^+$-$\Gamma_8^+$ (Fig. \ref{fig1}(f)). The predicted ordering changes into $\Gamma_7^+$-$\Gamma_7^-$-$\Gamma_8^+$-$L_6^+$ in DFT-HSE (Fig. \ref{fig1}(g)), which is consistent with experimental data~\cite{madelung2012}. For (111) films, the bulk $L$ pockets are projected onto $\overline{\Gamma}$ and $\overline{\textrm{M}}$ of the surface Brillouin zone (BZ) (Figs. \ref{fig1}(b) and (e)). With $L_6^+$ predicted below the Fermi level by DFT-PBE, the calculated (111) films are prone to be metallic\cite{chou2014}. This situation might be changed in DFT-HSE and in experiments, since $L_6^+$ is actually above the Fermi level in the bulk. For (100) and (110) films, the bulk $L$ pockets are projected onto $\overline{\textrm{X}}$/$\overline{\textrm{Y}}$ and $\overline{\textrm{X}}$/$\overline{\textrm{M}}$, respectively. Thus the near-Fermi-level $L$ and $\Gamma$ pockets do not interact with each other in these films. This feature simplifies the discussion of topological band inversions at $\overline{\Gamma}$ and is advantageous for opening large energy gaps in (100) and (110) films. A previous work calculated electronic structures of (100) films to study the 2D topological surface states of $\alpha$-Sn bulk\cite{kufner2014}, but the use of (100) films as 2D TIs has not been discussed before.

\begin{figure*}[htbp]
	\includegraphics[width=0.9\linewidth]{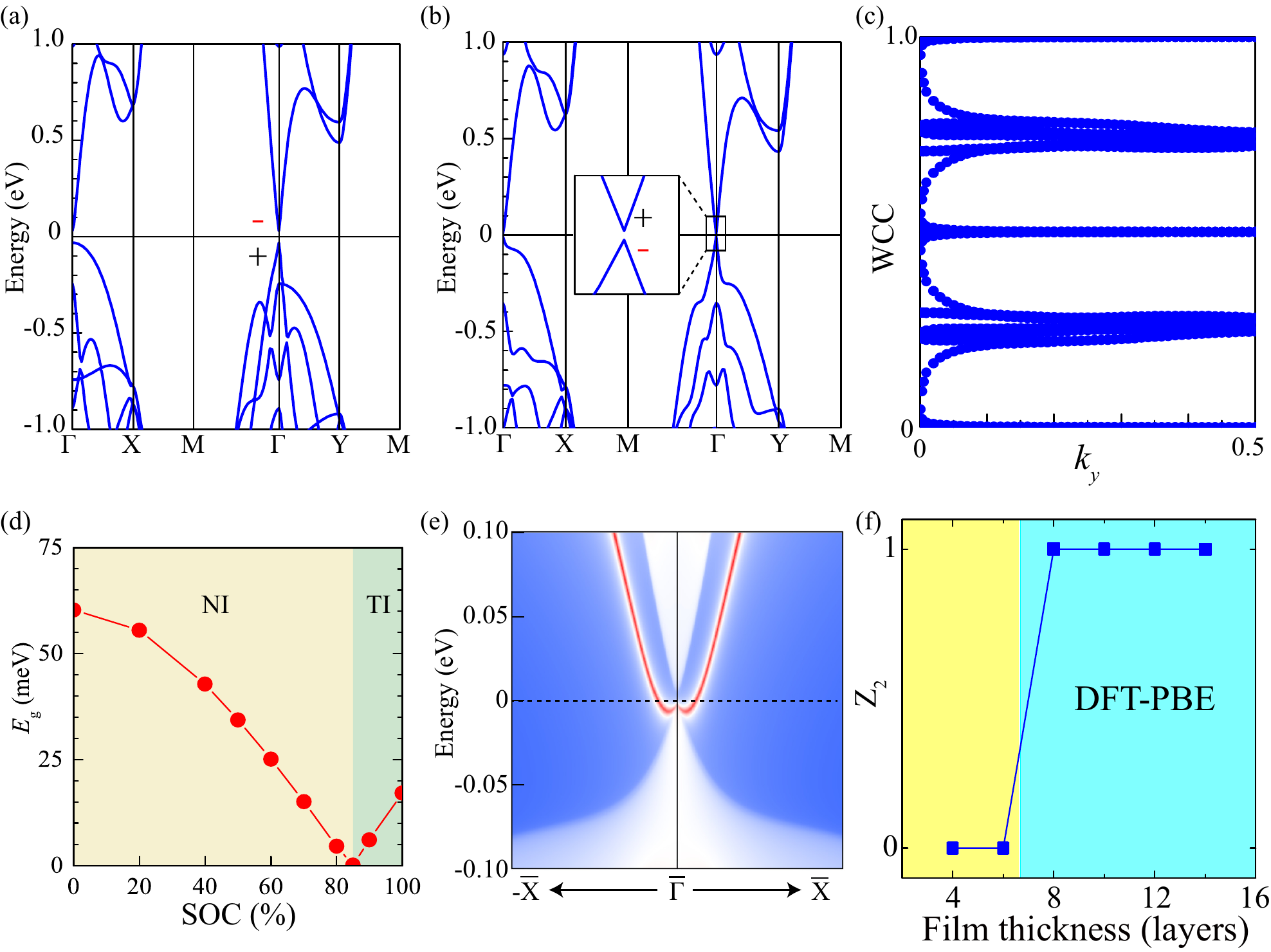}
	\centering
	\caption{Electronic structures and topological properties of $\alpha$-Sn(100) films calculated by DFT-PBE. (a,b) Band structures of an 8-layer $\alpha$-Sn (100) film (a) excluding and (b) including the SOC. Parities of Bloch wave functions at $\Gamma$ are labeled by “$+$” or “$–$”.  An enlarged band structure near $\Gamma$ is shown in the inset. (c) Evolution of Wannier charge centers (WCCs) in the Brillouin zone. (d) Direct band gap at $\Gamma$ as a function of the SOC strength. The system transits from the normal insulator (NI) phase into the TI phase. (e) Topological edge states of the semi-infinite film of 8-layer $\alpha$-Sn(100), displaying one pair of gapless helical edge states within the bulk gap. (f) Topological invariant $Z_2$ of $\alpha$-Sn(100) films as a function of film thickness.} \label{fig3} 
\end{figure*}

Figure \ref{fig2}(a) shows the optimized atomic structure of an 8-layer $\alpha$-Sn (100) film, whose dangling bonds are saturated by hydrogen. The equilibrium lattice constants are $a=4.75$ \AA, $b=4.69$ \AA, and the Sn-H bond length is 1.74 \AA. Due to the cubic symmetry, $a = b$ should be the case in the thick limit. Here the slight difference between $a$ and $b$ is caused by surface effects. We also calculated the phonon dispersions for the film. As shown in Fig. \ref{fig2}(c), no imaginary phonon mode appears (despite minor numerical inaccuracies), indicating that the film is dynamically stable and could be  synthesized experimentally.

\begin{figure*}[htbp]
	\includegraphics[width=0.7\linewidth]{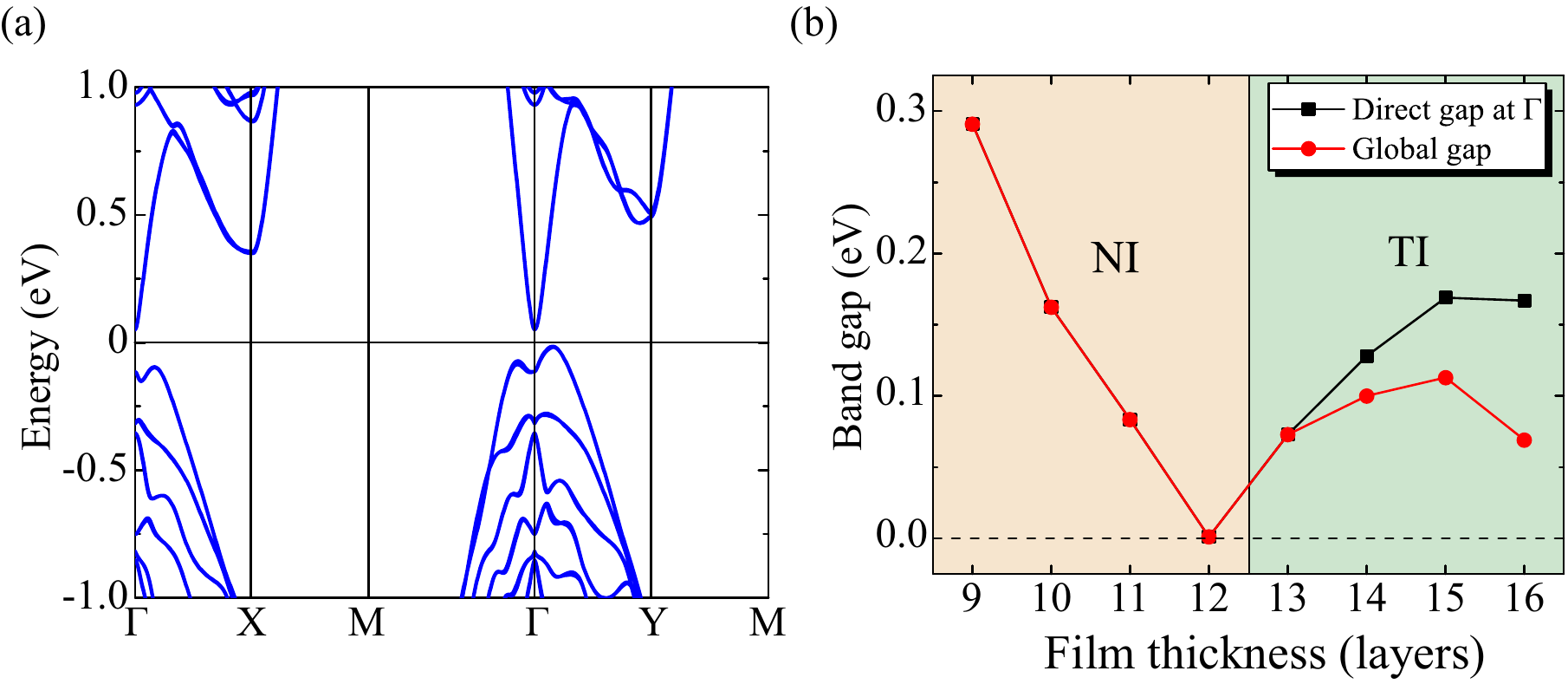}
	\centering
	\caption{Electronic structures and topological properties of $\alpha$-Sn(100) films calculated by DFT-HSE. (a) Band structure of a 16-layer $\alpha$-Sn (100) film including the SOC. (b) Direct band gap at $\Gamma$ and global band gap of $\alpha$-Sn(100) films as a function of film thickness. The NI and TI phases are characterized by $Z_2=0$ and $Z_2=1$, respectively, which are denoted by different colors.} \label{fig4} 
\end{figure*}

Figures \ref{fig3} (a) and (b) show the DFT-PBE band structures of 8-layer $\alpha$-Sn(100) film excluding and including the SOC, respectively. A topological invariant $Z_2$ is employed to identify the QSH insulator phase. $Z_2=1$ and $Z_2=0$ characterize topologically nontrivial and trivial phases, respectively. Due to the existence of inversion symmetry in the even-layer film, the topological invariant can be  calculated by the parity criterion\cite{fu2007}. The top valence bands and bottom conduction bands have opposite parities, and the SOC induces a band inversion between them (Fig. \ref{fig3} (d)), indicating an SOC-induced topological phase transition. Obviously, the gapped system with (without) the SOC has $Z_2=1$ ($Z_2=0$), suggesting that the 8-layer (100) film is a QSH insulator. The $Z_2$ value can also be determined through tracking the motion of Wannier charge centers of all occupied eigenstates\cite{gresch2017} or by performing edge-state calculations. These calculations consistently give $Z_2=1$ as shown in Figs. \ref{fig3} (c) and (e). Then we varied the film thickness, and calculated the thickness-dependent $Z_2$ as shown in Fig. \ref{fig3}(f). The results indicate that $Z_2 = 0$ for ultrathin films and $Z_2$ changes to 1 when the thickness is above a critical value. The critical thickness predicted by DFT-PBE is about 7-layer. However, since DFT-PBE is not able to accurately describe electronic structure of $\alpha$-Sn bulk, its quantitative predictions on the band gap values or even $Z_2$ values of thin films are questionable. For this reason, DFT-HSE calculations will be invoked in the following.

\begin{figure*}[htbp]
	\includegraphics[width=0.7\linewidth]{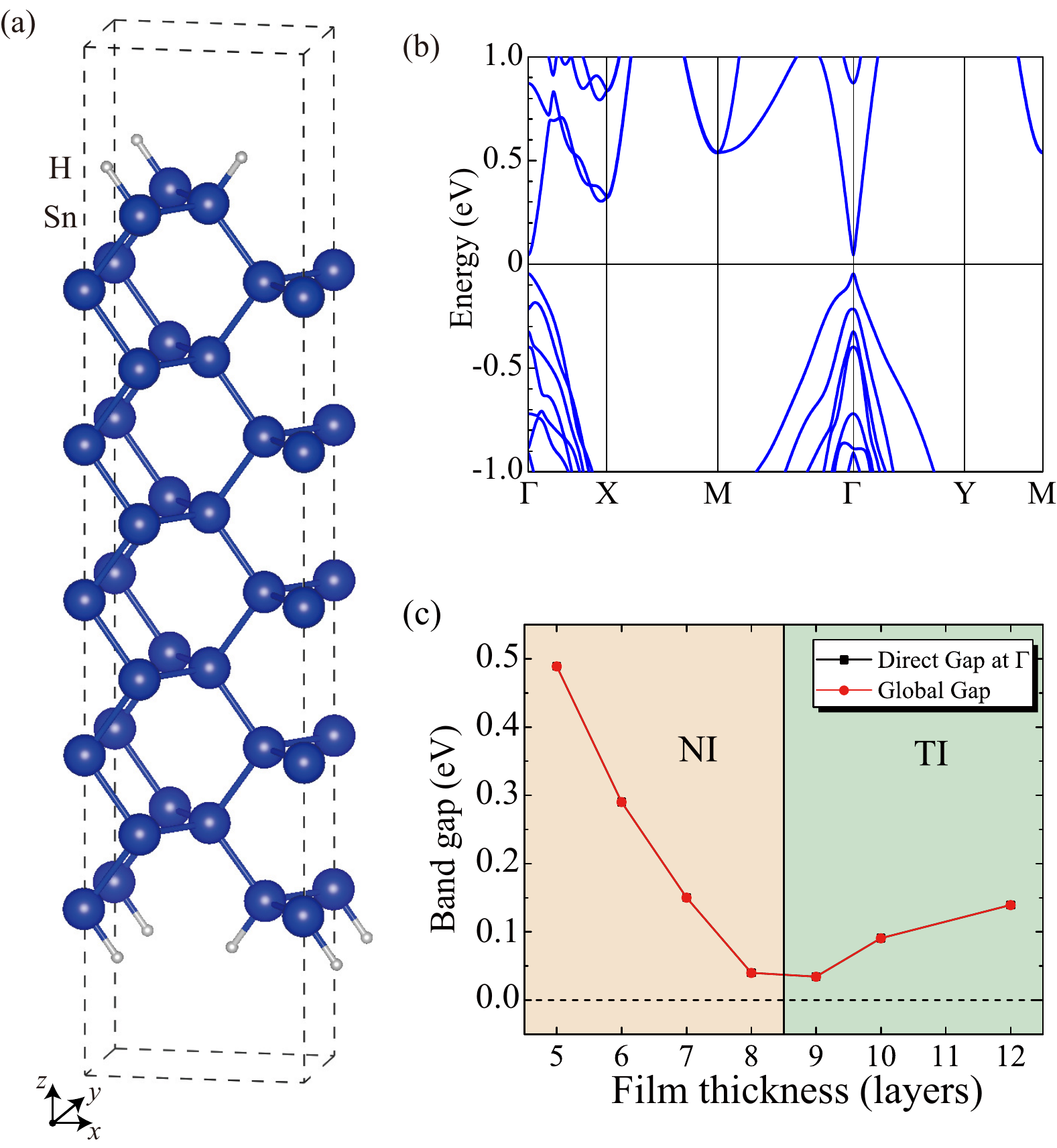}
	\centering
	\caption{Physical properties of $\alpha$-Sn (110) films, with electronic structures calculated by DFT-HSE. (a) The optimized atomic structure ($a=6.49$ \AA, $b=4.59$ \AA) and (b) band structure of a 10-layer $\alpha$-Sn (100) film including the SOC. Sn and H atoms are denoted by blue and gray balls, respectively. (c) Direct band gap at $\Gamma$ and global band gap of $\alpha$-Sn(110) films as a function of film thickness. The NI and TI phases are characterized by $Z_2=0$ and $Z_2=1$, respectively, which are denoted by different colors.} \label{fig5} 
\end{figure*}

Figure \ref{fig4}(a) displays the DFT-HSE band structures of 16-layer $\alpha$-Sn(100) film including the SOC. The ``M'' shape valence bands appear near the  
$\Gamma$ point, which is typically found for inverted band structures. Further analysis confirms that a topological band inversion exists in the film, and the system gives $Z_2=1$. In contrast to the result of DFT-PBE, $Z_2=0$ for 8-layer (100) film. Figure \ref{fig4}(b) presents the thickness-dependent band gap values and $Z_2$ values of $\alpha$-Sn(100) films predicted by DFT-HSE. The ultrathin films are semiconductors with direct band gaps at $\Gamma$. The band gap becomes indirect in thick films, due to the appearance of ``M'' shape bands. Usually, one would expect a monotonic decrease of band gap with increasing film thickness $d$, according to the fundamental physics of quantum confinement. However, the band gap here first decreases and then increases with increasing film thickness $d$. Limited by the computational resource, we can only compute films with $d < 16$ layers. Further increasing $d$, the band gap would decrease again. Eventually, the band gap would converge to zero (i.e., the bulk value) in the limit of infinite $d$. Therefore, the band gap as a function of $d$ shows an oscillating change behavior, which is an indicative of topological phase transitions\cite{liu2010}. Our $Z_2$ calculations indicate that $Z_2$ changes from 0 to 1 at a critical thickness of $d_0^{(1)} \sim 12$ layers.  $Z_2$ would change back to 0 at another critical thickness $d_0^{(2)} > d_0^{(1)}$. Obviously, $d_0^{(2)}$ is significantly larger than 16 layers. We thus have a wide thickness range $d_0^{(1)}<d<d_0^{(2)}$ to get QSH insulators. For (100) films from 13 to 16 layers, their topological energy gaps are as large as 0.1 eV.

Moreover, we performed similar calculations for $\alpha$-Sn(110) thin films. Using the 10-layer film as an example, the optimized atomic structure is displayed in Fig. \ref{fig5}(a), where the equilibrium lattice constants are $a=6.49$ \AA, $b=4.59$ \AA, and the Sn-H bond length is 1.74 \AA. Figure \ref{fig5}(b) presents its band structure calculated by DFT-HSE. An SOC-induced band inversion is still observed at the $\Gamma$ point, like in the (100) case. The thickness dependent properties of (110) films are summarized in Fig. \ref{fig5}(c). All the calculated thin films give direct band gaps at $\Gamma$. Similar as in (100) films, the band gap first decreases and then increases with increasing $d$. The largest $d$ we considered for (100) films is 12 layers. For thicker films, the band gap is expected to decrease when $d$ is large enough, and will finally reduces to zero in the bulk limit. Thus, an oscillating change behavior is also expected. Each critical thickness $d_0$ at which the band gap closes refers to a topological phase transition. For instance, the first critical thickness $d_0^{(1)}$ is between 8 and 9 layers, and $Z_2$ changes from 0 to 1 when $d$ increases across $d_0^{(1)}$. The second critical thickness $d_0^{(2)}$ should be significantly larger than 12 layers, and $Z_2$ varies from 1 to 0 when $d$ increases across $d_0^{(2)}$. Therefore, $Z_2 = 1$ for $d_0^{(2n-1)}<d<d_0^{(2n)}$ ($n=1,2,3...$), and  otherwise $Z_2 = 0$. Generally, the QSH gap with the largest energy gap will show up in the first window $d_0^{(1)}<d<d_0^{(2)}$. Indeed large-gap QSH insulators can be obtained in this thickness range. For instance, the topological energy gap is 0.14 eV for the 12-layer (110) film.

\section{Summary}
In brief, we studied thickness-dependent electronic and topological properties of $\alpha$-Sn(100) and $\alpha$-Sn(110) thin films by first-principles calculations. For both types of films, we predicted that the QSH phase exists in a wide thickness range and the QSH gap can be large ($>0.1$ eV). The finding suggests a new direction to search for QSH insulators beyond stanene in the $\alpha$-Sn system. Noticeably, a strict control of substrate and film thickness is not necessary in the present proposal, which facilitates the experimental realization. In fact, the growth of $\alpha$-Sn(100) and $\alpha$-Sn(110) thin films has been well demonstrated to be experimentally feasible on CdTe and InSb substrates\cite{barfuss2013,ohtsubo2013,farrow1981,tang1987,fantini2000,betti2002}. We thus believe that the large-gap QSH phase is very likely to be realized in the $\alpha$-Sn system. The same physics are applicable to other bulk materials with ``negative energy gaps'' (like HgTe\cite{madelung2012}, HgSe\cite{madelung2012} and Na$_3$Bi\cite{collins2018,xia2019}), where the QSH phase can be obtained by simply growing thin films in no need of building quantum well structures.

\section*{Acknowledgment}
This work is supported by the Basic Science Center Project of NSFC (Grant No. 51788104), the Ministry of Science and Technology of China (Grants No. 2018YFA0307100 and No. 2018YFA0305603), the National Natural Science Foundation of China (Grant No. 11874035) and the Beijing Advanced Innovation Center for Future Chip (ICFC).

\end{multicols}
\end{document}